\documentclass[pre,preprint,groupedaddress]{revtex4}

\usepackage{graphics}
\def\F{\Phi}
\def\k{\kappa}
\def\lm{f}
\def\l{\lambda}
\def\ord{{\cal O}}
\def\x{x}
\def\tf{\tilde{f}}
\def\BS{Bak-Sneppen }
\def\xor{\oplus}
\newcommand{\PP}{{\bf P }}
\newcommand{\NC}{{\bf NC }}

\begin{document}

\title{Parallel dynamics and computational complexity of the Bak-Sneppen model}
\author{J. Machta}
\email[]{machta@physics.umass.edu}
\author{X.-N. Li}
\address{Department of Physics,
University of Massachusetts,
Amherst, MA 01003-3720}
\begin{abstract}
The parallel computational complexity of the \BS evolution model is studied.  It is
shown that  \BS histories can be generated by a massively parallel
computer in a time that is polylogarithmic in the length of the history.  In this parallel
dynamics, histories are built up via a nested hierarchy of avalanches.  Stated in
another way, the main result is that the {\em logical depth} of producing a
\BS history is exponentially less than the length of the history.  This finding is surprising
because the self-organized critical state of the \BS model has long range correlations in
time and space that appear to imply that the dynamics is sequential and history
dependent.   The parallel dynamics for generating \BS histories is contrasted to standard
\BS dynamics.   Standard dynamics and an alternate method for generating histories,
conditional dynamics, are both shown to be related to P-complete natural decision
problems implying that they cannot be efficiently implemented in parallel.
\end{abstract} 


\maketitle


\section{Introduction}

The \BS model~\cite{BaSn,PaMaBa96}  is among the simplest systems to
display self-organized criticality and avalanche behavior.  Starting from arbitrary initial
conditions, the \BS model spontaneously reaches a state with long range correlations in
space and time that are set up by a series of avalanche events.  
The approach to criticality, the critical
state of the model, and the properties of avalanches in the \BS model have been
extensively studied during the past decade.  There are now precise values of critical
exponents and well-tested scaling laws.  On the other hand, there is not yet a theory of
the critical state except for the case of mean field theory and near zero dimension. 

In this paper we investigate the complexity of the histories generated by the \BS model. 
These histories manifest long time correlations with early random events influencing
distant, later events.   However, the past may determine the future in ways which are
more or less complex.   It is the purpose of this paper to give a precise meaning to the
notion of a complex history and then to show that histories produced by the \BS model
are not complex in this sense.  

The framework in which we will define the complexity of a history is computational
complexity theory.  Computational complexity theory measures the computational
resources needed to solve problems.  Among those resources are {\em parallel time} or
equivalently {\em depth}, which measures the number of steps needed by a massively
parallel computer to solve a problem.  If the solution to a problem can be re-organized
so that many computations are carried out simultaneously and independently then a
great deal of computational work can be accomplished in few parallel steps.  On the
other hand, it may be that the sequential nature of the problem precludes reorganizing
the work into a small number of parallel steps.  The theory of parallel computational
complexity provides algorithms for efficiently solving some problems in parallel and
provides evidence supporting the proposition that solutions to other tasks cannot be
substantially accelerated by parallelization.  A powerful feature of this theory is that it
yields essentially the same results  independent of the choice of the model of
computation.  Thus, measures of parallel computational complexity reveal the logical
structure of the problem at hand rather than the limitations of the particular model of
parallel computation.  

We will define the complexity of histories generated by a stochastic model such as
the \BS model in terms of the parallel time required to generate a sample
history.  If, through parallelization, the dynamics of the model can be reorganized so
that many processors running for a few steps can generate a statistically valid 
history then we will say the histories generated by the model are not
complex.   If on the other hand, there is little benefit from parallelization so that a
history must be generated in a step by step fashion, then the model generates complex
histories.  The  \BS model illustrates the point that scale invariant spatial and temporal
correlations are not sufficient to guarantee complexity in the sense defined here. 
Indeed, it is the strictly hierarchical nature of the avalanches in the \BS model that
permits us to generate them in a small number of highly parallel steps.  

The present work fits into a series of studies of the computational complexity of
models in statistical physics.  Although previous studies focused mainly on the
computational complexity of generating spatial patterns, in many cases the analysis
can also be applied to reveal the complexity of the history of the growth of the pattern. 
For example, two dynamical processes for diffusion limited aggregation (DLA)
have been shown to be P-complete~\cite{Mac93a,MaGr96}.  These results 
suggest that these processes for generating DLA clusters cannot be re-organized to
greatly reduce the parallel time needed for the generation of an aggregate (however, see
[\onlinecite{MoMaGr97}] for a parallel algorithm that achieves a modest speed-up for
DLA).   The growth of an aggregate can be viewed as a history and the same arguments
can be employed to give evidence that such histories cannot be significantly parallelized. 
Similarly, predicting the behavior of Abelian sandpiles for $d \geq 3$ has been shown to
be P-complete \cite{MoNi}.  By contrast,  the patterns generated by a number of growth
models such as the Eden model, invasion percolation and the restricted solid-on-solid
model were shown to require little parallel time~\cite{MaGr} and these results carry
over immediately to fast parallel algorithms for generating histories for these models.

\subsection*{Summary}
In Sec.\ \ref{sec:comp} we give an overview of selected aspects of computational
complexity theory relevant to the present work and in Sec.\ \ref{sec:bsmodel} we review
the \BS model.  In Sec.\ \ref{sec:parbs} we present parallel dynamics for the \BS model
and in Sec.\ \ref{sec:simpar} we present simulation results showing that  \BS
histories can be efficiently simulated using parallel dynamics.  In Sec.\
\ref{sec:bspc}  we show that the standard dynamics for the \BS model and a new
dynamics based on conditional probabilities both yield  P-complete problems.  The
paper concludes in Sec.\ \ref{sec:conc}.

\section{Computational complexity}
\label{sec:comp}

Computational complexity theory determines the scaling of computational resources
needed to solve problems as a function of the size of the problem.  Introductions to the
field can be found in Refs.\ [\onlinecite{Papa}]--[\onlinecite{GiRy}] .  
Computational complexity theory may be set up in several nearly equivalent ways
depending on the choice of model of computation.  Here we focus on parallel
computation and choose the standard {\em parallel random access machine} (PRAM) as
the model of computation.  The main resources of interest are {\em parallel time} or
{\em depth} and number of processors.  A PRAM consists of a number of simple
processors (random access machines or RAMs) all connected to a global memory. 
Although a RAM is typically defined with much less computational power than a real
microprocessor such as Pentium, it would not change the scaling found here to think of a
PRAM as being composed of many microprocessors all connected to the same random
access memory.   The processors run synchronously and each processor runs the same
program. Processors have an integer label so that different processors  follow different
computational paths.  Since all processor access the same memory, provision must be
made for how to handle conflicts.  Here we choose the {\em concurrent read, concurrent
write}  (CRCW) PRAM model in which many processors may attempt to write to or read
from the same memory cell at the same time.  If there are conflicts between what is to
be written, the processor with the smallest label succeeds in writing.    

The PRAM is an idealized and fully scalable model of computation.  The number of
processors and memory is allowed to increase {\em polynomially} (i.e.\ as an arbitrary
power) in the size of the problem to be solved.  Communication is idealized in that it is
assumed that any processor can communicate with any memory cell in a single time
step.  Obviously, this assumption runs up against speed of light or hardware density
limitations.  Nonetheless, parallel time on a PRAM quantifies a fundamental aspect of
computation.  During each parallel step, processor must carry out independent actions. 
Communication between processor can only occur from one step to the next via reading
and writing to memory.  Any problem that can be solved by a PRAM with $H$
processors in parallel time $T$ could also be solved by a single processor machine in a
time that is no greater than $HT$ since the single processor could sequentially run
through the tasks that were originally assigned to the
$H$ processors.  On the other hand, it is not obvious whether the work of a single
processor can be re-organized so that it can be accomplished in a substantially
smaller number of steps by many processor working independently during each step.  
Two examples will help illustrate this point.  First, suppose the task is to add
$n$ finite-precision numbers.  This can be done by a single processor in a time that
scales linearly in $n$.  On a PRAM with $n/2$ processor this can be done in $\ord(\log
n)$ parallel time using a binary tree (for simplicity, suppose $n$ is a power of $2$).  In
the first step, processor one adds the first and second numbers and puts the
result in memory, processor two adds the third and fourth numbers and puts the result
in memory and so on.  After the first step is concluded there are $n/2$ numbers to add
and these are again summed in a pairwise fashion by
$n/4$ processors.  The summation is completed after $\ord(\log n)$ steps.   This
method is rather general and applies to any associative binary operation.   Problems of
this type have efficient parallel algorithms and can be solved in time that is a power of
the logarithm of the problem size, here $n$, that is,  {\em polylog} time.

A second example is the problem of evaluating the output of a Boolean circuit
with definite inputs.  A Boolean circuit is composed of AND, OR and NOT gates
connected by wires.  Suppose the gates are arranged in levels so
that gates in one level take their inputs only from gates of the previous level and give
their outputs only to gates of the next level so there is no feedback in the circuit.  The
number of levels in the circuit is referred to as the {\em depth} of the circuit.  At the top
level of the circuit  are the TRUE or FALSE inputs.   At the bottom level of the
circuit are one or more outputs. Given some concisely encoded description of the circuit
and its inputs,  the problem of obtaining the outputs is known as the {\em circuit value
problem} (CVP). Clearly, we can solve CVP on a  PRAM in parallel time that is
proportional to the depth of the circuit since we can evaluate a single level of the circuit
in a single parallel time step.  On the other, there is no known general way of speeding
up the evaluation of a Boolean circuit to a time that is, say polylog in the depth of the
circuit. For the case of adding $n$ numbers, the logical structure of the problem is
sufficiently simple that we can substitute hardware for time to obtain the answer in
polylog parallel steps.  For CVP,  hardware cannot be used to substantially decrease the
depth of the problem.

At the present time, the statement that CVP cannot be solved in polylog parallel time is
not mathematically proved.  Instead, one can show that CVP is {\em P-complete}.  To
understand the meaning of P-completeness we must first introduce the complexity
classes \PP and \NC and the notion of {\em reduction}.  \PP consists of the class of
problems that can be solved by a PRAM in polynomial time with polynomially many
processors.  \NC consists of the class of problems that can be solved in polylog time on a
PRAM with polynomially many processors.  A problem ${\cal A}$ is reduced to a
problem ${\cal B}$ if a PRAM with an oracle for ${\cal B}$ can be used to solve ${\cal
A}$ in polylog time with polynomially many processors.  An oracle for ${\cal B}$ is able
to supply the PRAM a solution to an instance of ${\cal B}$ in a single time step.
Intuitively, if ${\cal A}$ can be reduced to ${\cal B}$ then ${\cal A}$ is no harder to
solve in parallel than
${\cal B}$.  A problem in \PP is P-complete if all other problems in \PP can be reduced
to it.  It can be proved that CVP is P-complete.  Furthermore, if it could be shown that
there is any problem in \PP that is not in \NC then it would follow that CVP is not in
\NC.  However, the conjecture that $\PP\neq\NC$, though widely believed, is not yet
proved.  Since reductions are transitive, showing that another problem ${\cal A}$ is 
P-complete can  be proved by reducing CVP to ${\cal A}$.  Later in the paper we will
take this approach to prove that two problems representing different \BS dynamics are
P-complete.

In addition to supplying a P-complete problem, Boolean circuits serve as an alternate
model of computation.  For a given computational problem, 
and for a given problem, there is a Boolean circuit that solves the problem.  In
order to solve problems of any size, a family of Boolean circuits is needed, one for each
problem size.  Furthermore,  to be comparable to a PRAM running a single program, the
family of circuits should satisfy a {\em uniformity} condition that bounds the
complexity of the computation needed to specify each circuit in the family.  The
resources of circuit depth (number of levels in the circuit) and circuit width (maximum
number of gates in a level) are roughly comparable to the PRAM resources of parallel
time and number of processor, respectively.  Indeed the class \NC can be defined as the
set of problems that can be solved by uniform circuit families whose depth scales
polylogarithmically in the problem size and whose width scales polynomially in the
problem size.  \PP is the class of problems solvable by circuits of polynomial width and
depth.

\section{\BS model}
\label{sec:bsmodel}
In this section we review the \BS model and its behavior.  A detailed
discussion can be found in Ref.\ \onlinecite{PaMaBa96}. The \BS model
defines histories of configurations on a
$d$-dimensional lattice.  At each lattice site $i$ and discrete time $t$, there is a number
$\x_i(t) \in [0,1]$.  The {\em
standard} dynamics for the \BS model creates a history sequentially with one new
time slice determined in each step of the dynamics.  A single time step in  standard
dynamics for the \BS model consists of locating the {\em extremal}  site $s$,
which has the minimum number.   The extremal site and its
$2d$ nearest neighbors are given new $\x$'s chosen randomly from the uniform
distribution on
$[0,1]$.  The $\x$'s on the remaining sites are unchanged. Let
$\lm(t)$ be the 
extremal value at the extremal site $s(t)$ at time $t$.  

\BS histories can be organized into {\em avalanches} or
causally connected regions of activity in the $d+1$-dimensional space-time.  Avalanches
can be classified by bounding the extremal value during the avalanche.  The collection of
renewed sites from $t$ to 
$t^\prime$ is an $\tf$-avalanche if $\lm(t-1) \geq \tf$ and $\lm(t^\prime) \geq \tf$ but
for all
$t^{\prime\prime}$ with $t \leq t^{\prime\prime} <t^\prime$, $\lm(t^{\prime\prime}) <
\tf$. Note that it is also possible to have an $\tf$-avalanche whose duration is a single
time step (i.\ e.\ $t=t^{\prime}$ and no times $t^{\prime\prime}$) if $\lm(t-1) \geq \tf$
and $\lm(t) \geq \tf$.  An $\tf$-avalanche can be partitioned  into a sequence of one or
more $\tf^\prime$-avalanches for 
$\tf^\prime < \tf$.  Repeated use of this idea allows us to partition avalanches into a
nested sequence of subavalanches. The partitioning of avalanches into subavalanches is
one of the key ideas in the parallel algorithm described
below.

\begin{figure}[h]
\includegraphics{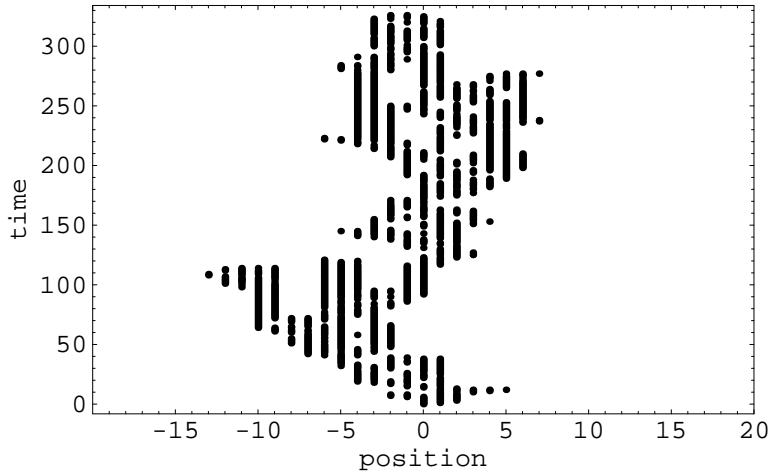}
\caption{The backbone of an $\tf$-avalanche for $\tf=0.5$  in the one-dimensional \BS
model. Black dots represent positions where the random numbers are less than $\tf$. 
Immediately before and at the last time of the avalanche all random numbers are
greater than $\tf$.}
\label{fig:f13}
\end{figure}

The properties of avalanches are entirely independent of the environment
in which they exist so long as the environment is large enough to contain the
avalanche.  Indeed we can grow the {\em backbone} of an avalanche without regard
to the environment.  The backbone of an $\tf$-avalanche is the set of sites $i$ and
the corresponding $\x_i$ for which $\x_i < \tf$. (Note that the last time step in an
avalanche is not part of the backbone).  The backbone of an
$\tf$-avalanche  for
$\tf=0.5$ is shown in Fig.\ \ref{fig:f13}.  The only information we need about the
environment is that all environmental sites have random numbers uniformly distributed
in the interval $[\tf,1]$.  The values of these numbers is unimportant and the full lattice
of random numbers does not have to be realized.   During the construction of an
$\tf$-avalanche, sites must be renewed.  Instead of assigning random numbers in the
interval $[0,1]$ as is done in standard \BS dynamics, one can declare a renewed site to
be part of the environment with probability
$1-\tf$ or, with probability $\tf$, declare it part of the backbone.  If the site is part of
the backbone it is then given an explicit random number in the interval 
$[0,\tf)$.  The construction of an $\tf$-avalanche backbone begins with the renewal of
the origin and its neighbors and then continues until there are no remaining sites with
number less than $\tf$.  This approach for making an avalanche backbone is called the
\BS branching process in [\onlinecite{PaMaBa94}] and is an efficient method for carrying
out large-scale simulations of the \BS model.  

The \BS model manifests self-organized criticality: independent of the initial
state,  after a long time the system approaches a stationary state
where the average avalanche size diverges and the distribution of avalanche sizes
follows a power law.  In the critical state almost all $\x$'s exceed a critical value,
$f_c$ (for the $d=1$, $f_c\approx 0.6670$).  If $S$ is the duration
of an avalanche then, in the critical state, the probability of an avalanche of duration $S$
decays as $S^{-\tau}$ ($\tau\approx1.1$ for $d=1$).  The average duration $\langle S
\rangle_{\tf}$ of $\tf$-avalanches diverges as
\begin{equation}
\label{eq:gamma}
\langle S
\rangle_{\tf} \sim (f_c - \tf)^{-\gamma}
\end{equation}
with $\gamma\approx 2.7$ for $d=1$.  The
$\tf$-avalanche size distribution, ${\cal P}(S,\tf)$  has the scaling form ${\cal
P}(S,\tf)=S^{-\tau} g(S(f_c - \tf)^{1/\sigma})$ with $\sigma$, $\tau$ and $\gamma$
related by the usual exponent relation $\gamma= (2-\tau)/\sigma$.  The scaling
function $g(y)$ goes to a constant as $y \rightarrow 0$ and decays rapidly to zero for $y
\gg 1$.  Thus the upper cut-off $S_{\rm co}$ in the $\tf$-avalanche size distribution 
obeys the scaling law
$S_{\rm co} \sim (f_c-\tf)^{1/\sigma}$ ($\sigma\approx 0.34$ for $d=1$ \cite{Grass95}). 
The spatial extent of $\tf$-avalanches also diverges in the critical state.  Let
$W$ be the number of sites covered by an $\tf$-avalanche, then $\langle W
\rangle_{\tf} \equiv (f_c-\tf)^{-1}$.

The approach to the critical state is characterized by a {\em gap} $G(t)$, which is a
stepwise increasing function of time and is the maximum value of
$f(t^\prime)$ for $t^\prime \leq t$.  Times when the gap increases  
mark the end of an avalanche.  If  the gap
first has the value $G$  at time $t$ then  all
$\x$'s in the system are greater than or equal to $G$ at time $t$ and a $G$-avalanche
begins at time
$t+1$.  On average, the gap approaches the critical value as a power law, 
\begin{equation}
\label{eq:gapt}
f_c- \langle G(t) \rangle \sim
(t/N)^{\frac{1}{\gamma-1}}
\end{equation}
where $N$ is the number of sites in the system. In the
parallel dynamics for the
\BS model we will first identify the successive values of the gap and then fill in the
intervals between them with the appropriate avalanches.
  
\section{Parallel \BS dynamics}
\label{sec:parbs}

In this section we present a parallel approach for constructing \BS histories.  At the core
of this approach is a procedure for hierarchically building avalanches in which
small avalanches are pasted together to yield larger avalanches.  The large scale
structure of the history is controlled by the sequence of gap values.  Each value of the
gap determines an avalanche that needs to be constructed.  The algorithm proceeds
in three stages.  In the first stage, described in Sec.\ \ref{sec:gap}, the gap sequence is
obtained and in the second stage, described in Sec.\ \ref{sec:parf}, the required
avalanches are constructed.  Finally, in the third stage, all space-time points that are not
yet determined are explicitly given numbers.

Although we have not run the parallel algorithm on a parallel computer, we have
measured essential aspects of its behavior using a conventional \BS simulation and the
results are reported in Sec.\ \ref{sec:simpar}.   The key finding is that we can produce
\BS histories in a parallel time that is polylogarithmic in the length of the
history.

\subsection{Parallel algorithm for the gap sequence}
\label{sec:gap}
In this subsection we show how to construct a sequence of gap values, $G_k$,
$k=1,\ldots, K$ for a system with $N$ lattice sites (independent of dimension).  Suppose
that at some time the gap increases from
$G_k$ to
$G_{k+1}$.  This event occurs at the last step of a $G_k$-avalanche so, before the
selection of the extremal value that increases the gap, the distribution on every site of
the lattice is uniform on the interval
$[G_k,1]$.  Thus
$G_{k+1}$ is a random variable given by
\begin{equation}
\label{eq:gap}
G_{k+1}= G_k + X (1-G_k) \equiv G_k \xor X
\end{equation}
where $X$ is a random variable that is the minimum over $N$ random numbers
uniformly distributed on $[0,1]$ and $\xor$ is the XOR operation for real
numbers.  Since $\xor$ is associative we can unroll  Eq.\
(\ref{eq:gap}) to obtain,
\begin{equation}
G_k = G_0 \xor X_1 \xor X_2 \xor \cdots \xor X_k
\end{equation}
where $G_0$ is the initial value of the gap and the $X_j$ are identically distributed
independent random variables chosen from the same distribution as $X$ described
above.  Using a standard parallel minimum program, each $X_j$ can be generated on a
probabilistic PRAM in $\ord(\log N)$ parallel time.   A standard parallel prefix
algorithm
\cite{GiRy} can then be used to evaluate all $K$ values of the gap in $\ord(\log K + \log
N)$ parallel time. 

\subsection{Parallel algorithm for $\tf$-avalanches}
\label{sec:parf}
In this section we show how to generate an $\tf$-avalanche in parallel time
that is polylog in $(f_c-\tf)$.  The idea of this algorithm is to generate in parallel a large
number of independent small avalanches and then to paste groups of these together to
create a smaller number of larger avalanches.  These are in turn pasted together to
create yet larger avalanches and so on in a hierarchical fashion.  The algorithm is
controlled by a pre-determined list of $f$ values, $f_0 < f_1 < \ldots < \tf$. These values
specify the types of avalanches to be manufactured. The sequence of $f$ converges to
$f_c$ as
\begin{equation}
\label{eq:fn}
f_n = f_c(1-q^n)
\end{equation}
with $q$ a constant between $0$ to $1$.    This choice for ${f_n}$ implies that for large
$n$, $\langle S
\rangle_{f_{n}}/\langle S \rangle_{f_{n-1}}
\rightarrow q^{-\gamma} $ so that the average number of
$f_{n-1}$-avalanches needed to construct an $f_n$-avalanche is asymptotically
constant (and equal to $q^{-\gamma}$).  The length of the sequence  and its penultimate
element 
$f_{n_{\rm max}}$ is chosen so that
$f_{n_{\rm max}}\leq\tf<f_{n_{\rm max}+1}$.  

Since $f_0=0$,  $f_0$-avalanches always have
$S=1$ and consist of the origin and its $2d$ nearest neighbors together with their
associated $\x$ values.  In parallel we can make $M$, $f_0$-avalanches in constant time
using
$M$ processors.  We now construct $f_1$-avalanches by pasting together one or more
$f_0$-avalanches.  The origin of the first $f_0$-avalanche is also the origin of the
$f_1$-avalanche. If the minimum of the $2d+1$ $\x$'s of the $f_0$-avalanche is greater
than $f_1$ then we are done and the $f_0$-avalanche is also an $f_1$-avalanche.  If not,
a second
$f_0$-avalanche is attached to the first $f_0$-avalanche with its origin lined up with the
extremal site of the first avalanche.  Again, we check whether the minimum $\x$ among
all the sites covered by the pair of $f_0$-avalanches is greater than $f_1$ and, if so, we
are done.   The pattern for constructing $f_{n}$-avalanches from an
$f_{n-1}$-avalanche should now be clear:
$f_{n-1}$-avalanches are pasted together until all the covered sites have
$\x \geq f_{n}$.  

The process of pasting together $f_{n-1}$-avalanches is a generalized version of the
standard \BS process.  Indeed, pasting together $f_0$-avalanches is exactly the standard
\BS process.   All that we need to know
about each $f_{n-1}$-avalanche is its origin and the final values of $\x$ on the sites
the avalanche covers.   During the construction of an $f_{n}$-avalanche, when a new
$f_{n-1}$-avalanche is attached, its origin is lined up with the extremal site among all
of the sites currently covered by the growing $f_{n}$-avalanche.  The $\x$ values on the
covered sites of the newly added
$f_{n-1}$-avalanche  renew  the sites covered by this avalanche and, if some of these
sites were previously part of the environment, the total number of covered sites of the
growing
$f_n$-avalanche is increased.  If the covered sites all have $\x \geq f_n$, the
$f_n$-avalanche is finished.  The pasting together of
$f_{n-1}$-avalanches must be carried out sequentially so that if the number of  
$f_{n-1}$-avalanches  needed to construct a single $f_{n}$-avalanches is $\kappa_n$ 
then it will take
$\kappa_n$ parallel steps to grow the avalanche.  Finally, once an $\tf$-avalanche is
constructed, it is straightforward to find its backbone by eliminating those sites for
which $\x_i \geq \tf$.  

On average $q^{-\gamma}$ 
$f_{n-1}$-avalanches must be pasted together to form one $f_{n}$-avalanche,  that is
$\langle \kappa_n \rangle =  q^{-\gamma}$.  However,  the $n^{\rm th}$ step in the
parallel construction of an $\tf$-avalanche is not complete until {\em all} the required 
$f_{n}$-avalanches have been made. Thus, the time required for this step will be set by
the maximum value of
$\kappa_n$.  As $\tf$ approaches
$f_c$, and for fixed
$n$ we need to make increasingly many $f_{n}$-avalanches so that the tail of the
$\kappa_n$ distribution is explored.  As we
shall see in the next section, the expected value of the maximum of $\kappa_n$
increases logarithmically in the number of $f_{n+1}$-avalanches that are constructed. 
As a result, the average parallel time required to generate an $\tf$-avalanche 
is predicted to be proportional to $\log^2(f_c-\tf)$. 

The discussion thus far has posited an effectively infinite environment for the avalanche
to grow in, however this is not required.  The same hierarchical construction can just as
well be used to make avalanches in periodic boundary conditions in a system of size
$L$.  Additional complications arise for other
boundary conditions due to the breaking of translational symmetry and this situation
will not be discussed here.  

Finally, we are ready to describe how to construct a full \BS history in a system with
$N$ sites and periodic boundary conditions starting from the  initial condition, $G_0=0$. 
First, following the procedure of Sec.\ \ref{sec:gap} we construct a 
sequence of $K$ gap values, $G_0, G_1,\ldots,G_K$.  Then, using the parallel algorithm
described above, for each gap value, $G_k$, we construct the backbone of a
$G_k$-avalanche for all $k <K$.  The site $i$ where $x_i=G_k$ is chosen randomly from
among all the sites of the lattice and this site serves as the spatial origin for the
$G_k$-avalanche.  The starting time $t_k+1$ of the $G_k$-avalanche is obtained by
summing over the durations of the preceding avalanches,
$t_k=\sum_{l=0}^k S_l$ where $S_l$ is the duration of the $G_l$-avalanche.  
Successive avalanche backbones are concatenated to form the backbone of the  history
which includes all the avalanche backbones and the intervening extremal sites that have
the gap values. 

Thus far only the backbone of the history is specified while sites in the environment
are conditioned by the extremal values but not fully specified.  A definite, statistically
valid, history is constructed from the backbone by fixing all space-time points
$x_i(t)$ not in the backbone. At the last time, $t_{\rm max}$, the gap
has the values $G_K$.  The randomly chosen extremal site, $s(t_{\rm max})$ is part of
the backbone and takes the value $G_K$.   All other sites at time $t_{\rm max}$ are
chosen from the uniform distribution on $[G_K,1]$. From these values we now work
backwards in time independently for each site.  Consider site $i$, we have that
$\x_i(t)=\x_i(t_{\rm max})$ until the most recent time that $i$ was part of the
backbone.   During the time intervals that site $i$ is part of the backbone, its value is
determined.  For each connected time interval,
$[t^\prime,t^{\prime \prime}]$ that site $i$ is part of the environment, $\x_i$ takes a
single value.  There are two possibilities.  The interval may come to end because $i$ is 
an extremal site where the gap increases from $G_{k-1}$ to $G_k$  that is, $t^{\prime
\prime}=t_k$ and $i=s(t_k)$.  In this case, $x_i=G_k$ throughout the interval.  The
second possibility is that site $i$ is renewed  at time $t^{\prime \prime}+1$ and
becomes part of the
$G_k$-avalanche backbone at that time.  In this case,  $x_i$ is chosen
uniformly from the distribution $[G_k,1]$ and has this value during the interval from
$t^{\prime}$ to $t^{\prime \prime}$.  Working backwards in this way a definite \BS
history is re-constructed from the backbone.  It should be noted that the construction
of a definite  history from the backbone can be carried out in parallel in polylog time
since each site can be treated independently and since intervals during which a site is
not in the backbone can be identified using a parallel graph connectivity algorithm.

\section{Simulation of  parallel dynamics}
\label{sec:simpar}
We have studied the performance of the parallel algorithm for constructing
avalanches in the one-dimensional \BS model.  We collected statistics on $\kappa_n$, the
number of
$f_{n-1}$-avalanches needed in the construction of  $f_{n}$-avalanches during the
manufacture of a single $f_{n_{\rm max}}$-avalanche.  Let $\Omega_n(n_{\rm
max})\equiv\max\{\kappa_n\}$ be the maximum value of $\kappa_n$ required during
the construction of the $f_{n_{\rm max}}$-avalanche.   The parallel running time
$T(n_{\rm max})$ , up to a constant of proportionality, for constructing an
$f_{n_{\rm max}}$-avalanche is obtained by summing $\Omega_n(n_{\rm
max})$,
\begin{equation}
\label{eq:tpar}
T(n_{\rm max})=\sum_{n=1}^{n_{\rm max}}  \Omega_n(n_{\rm max}) .
\end{equation}
The simulation works by running standard \BS dynamics and, on the fly, keeping track
of $\kappa_n$ and $\Omega_n$.

The following pseudo-code sketches a single step of an algorithm that collects statistics
on the distribution of $\kappa_n$ and computes the running time for constructing 
$f_{n_{\rm max}}$-avalanches.  It is based on the observation that an
$f_n$-avalanche is concluded when the extremal value $f(t) \geq f_n$.  Arrays
$\kappa_n$ and $\Omega_n$ are initialized to zero.  

\begin{enumerate}
\item Execute one time step of standard \BS dynamics and find the extremal value,
$f(t)$.
\item \textbf{For} $n=1$ to $n=n_{\rm max}-1$ \textbf{do}

      \textbf{if} $f(t)>f_{n}$ \textbf{then}
      \begin{itemize}
               \item $\kappa_{n+1} \leftarrow \kappa_{n+1}+1 $
               \item \textbf{if} $\kappa_n>\Omega_n$  \textbf{then}
                     $  \Omega_n \leftarrow \kappa_n $
								\item  Collect statistics for $\kappa_n$.
               \item $ \kappa_n \leftarrow 0 $
      \end{itemize}

      \textbf{end do}
\item \textbf{if} $f(t)>f_{n_{\rm max}}$  \textbf{then}  
      the $f_{n_{\rm max}}$-avalanche ends.
      Calculate the parallel time, $T$ by summing $\Omega_n$. 

      \textbf{else} go to Step 1.
\end{enumerate}

\begin{figure}[h]
\includegraphics{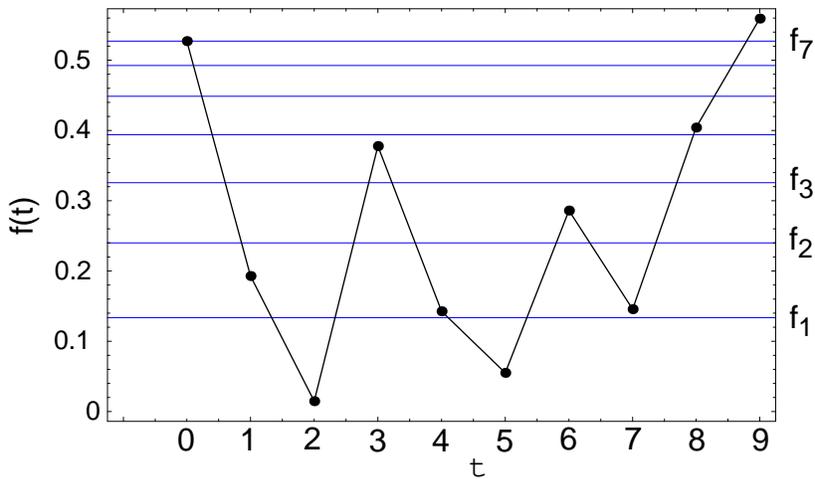}
\caption{Extremal values, $f(t)$ vs.\ time $t$ during an $f_7$-avalanche.  Horizontal
lines are the values
$f_1, \ldots, f_7$.}
\label{fig:min}
\end{figure}

\begin{figure}[!] 
\includegraphics{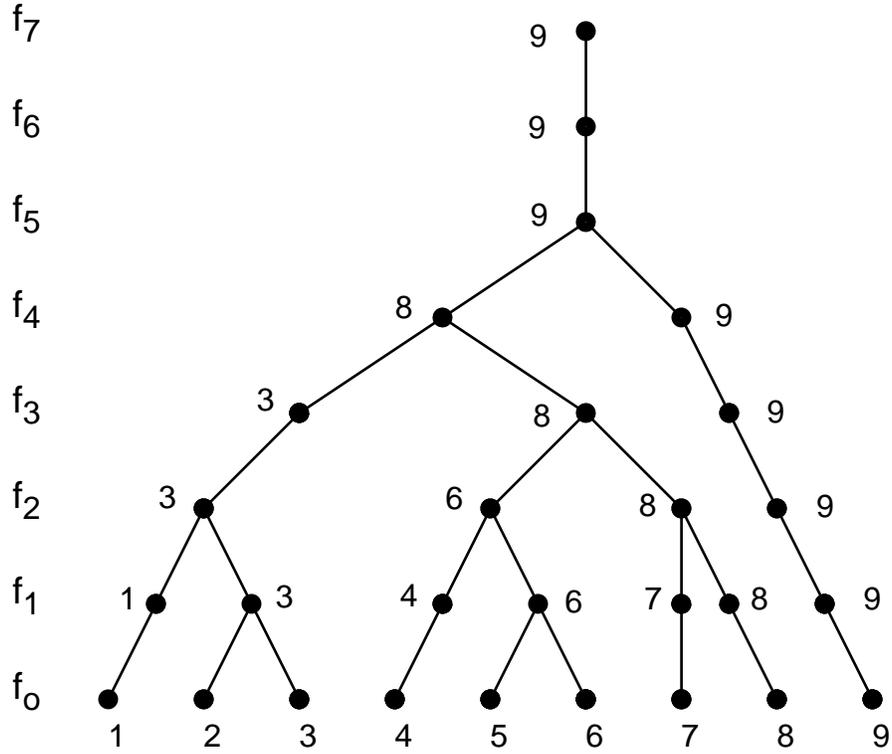}
\caption{Tree representing the re-organization of the $f_7$-avalanche of Fig.\
\ref{fig:min} into subavalanches.  The $f_n$-level of the tree represents partitioning the
$f_7$-avalanche into $f_n$-avalanches and the numbers on the nodes represent the
ending times of the $f_n$-avalanches. Roughly speaking, sequential time runs to the right
and parallel time runs upward.}
\label{fig:fseven}
\end{figure}

The re-organization of the sequential history of an avalanche into a hierarchy of
subavalanches is illustrated in Figs.\ \ref{fig:min} and \ref{fig:fseven} for the case of an
$f_7$-avalanche.  In Fig.\
\ref{fig:min} the sequence of extremal values is shown together with horizontal lines
indicating bounds $f_1, \ldots, f_7$.  Figure \ref{fig:fseven} shows a tree
representing the hierarchy of subavalanches making up the $f_7$ avalanche. 
The  $f_n$ level of the tree represents the $f_n$-avalanches needed in the
construction of the final $f_7$-avalanche.   The number next to each node represents
the ending time of a subavalanche.   For example, the
$f_3$ level of the tree shows that the
$f_7$-avalanche is composed of three $f_3$-avalanches ending at times 3, 8 and 9. The
total parallel time is the sum of the maximum degree of the tree at each level and is 13
in this case.

\begin{figure} 
\includegraphics{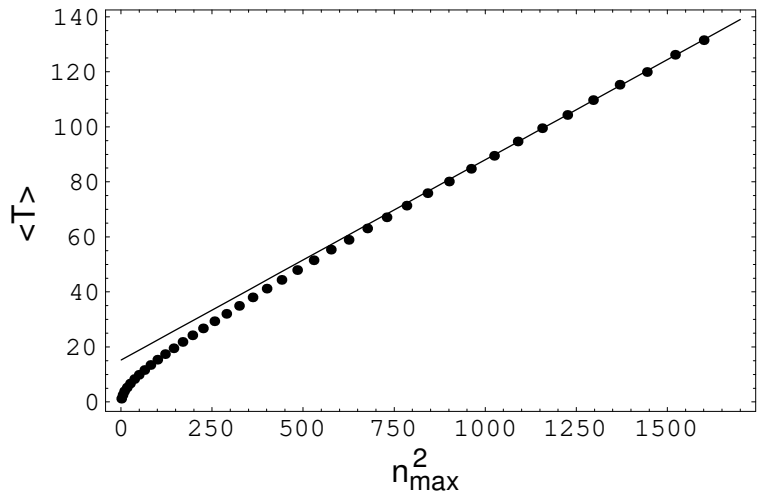}
\caption{The average parallel time $\langle T(n_{\rm max}) \rangle$
for constructing an $f_{n_{\rm max}}$-avalanche vs.\  $n_{\rm max}^2$. The straight
line is a linear fit to the data for $n_{\rm max} \geq 32$.} 
\label{fig:time}
\end{figure}

 In our simulations, for each $n_{\rm max}$ up
to 40 we constructed 
$M=10^{5}$  $f_{n_{\rm max}}$-avalanches.  The sequence of avalanche sizes
is given by Eq.\ \ref{eq:fn} with $q=0.9$ and $f_c=0.667$.
Figure \ref{fig:time} shows the $\langle T(n_{\rm max}) \rangle$ vs.\ $n_{\rm
max}^2$.  A fit of the data  to the functional form $\langle
T(n_{\rm max})
\rangle = a n_{\rm max}^z + c$ over the range $13\leq n_{\rm max} \leq 40$ yields
$z=1.832 \pm 0.004$.  The fit is reasonable with $\chi^2/{\rm d.o.f} = 1.1$.  We also
collected data  for
$q=0.8$ and $q=0.7$ over a smaller range of $n_{\rm max}$.  For $q=0.8$ we found that
$z=2.00
\pm 0.03$ ($\chi^2/{\rm d.o.f} = 1.3$) for
$12\leq n_{\rm max} \leq 20$ while for $q=0.7$ we found that $z=1.88 \pm 0.09$
($\chi^2/{\rm d.o.f} = 1.3$) for
$9\leq n_{\rm max} \leq 13$.  Since $\log(f_c-f_{n})
\sim n
\log q$, a conservative conclusion from the data is that the average parallel time
$\langle T
\rangle$ for constructing an
$\tf$-avalanche behaves as
$\langle T \rangle=\log^{\ord(1)} (f_c-\tf)$ or, in terms of the duration of the
avalanche$\langle T \rangle=\log^{\ord(1)}S$.

The time requirements of the parallel algorithm can be understood by
considering the probability density, $P_n(\kappa_n)$ of $\kappa_n$.  Recall that
$\kappa_n$ is the number of $f_{n-1}$-avalanches needed in the construction of an
$f_n$-avalanche and that its average is
$\langle
\kappa_n \rangle = q^{-\gamma}$.  Suppose that the tail of $P_n(\kappa_n)$ behaves as
$e^{-b \kappa_n}$.  Given these assumptions we can estimate the maximum value,
$\Omega_n(n_{\rm max})$, of
$\kappa_n$ in the construction of an $f_{n_{\rm max}}$-avalanche.
The estimate is obtained by requiring that the expected number of $f_n$-avalanches
manufactured times
$P_n(\Omega_n(n_{\rm max}))$ is order unity.  The expected number of
$f_n$-avalanches made during the construction of an $f_{n_{\rm max}}$-avalanche is
$q^{-(n_{\rm max}-n)\gamma}$ so we have
\begin{equation}
 \Omega_n(n_{\rm max}) \approx \frac{(n_{\rm max}-n) \ln(1/q^\gamma)}{b}
\end{equation}
Summing over $n$ yields
\begin{equation}
T(n_{\rm max}) \approx \frac{n_{\rm max}^2 \ln(1/q^\gamma)}{2b}
\end{equation}
in qualitative agreement with the numerical results.
 
\begin{figure} 
\includegraphics{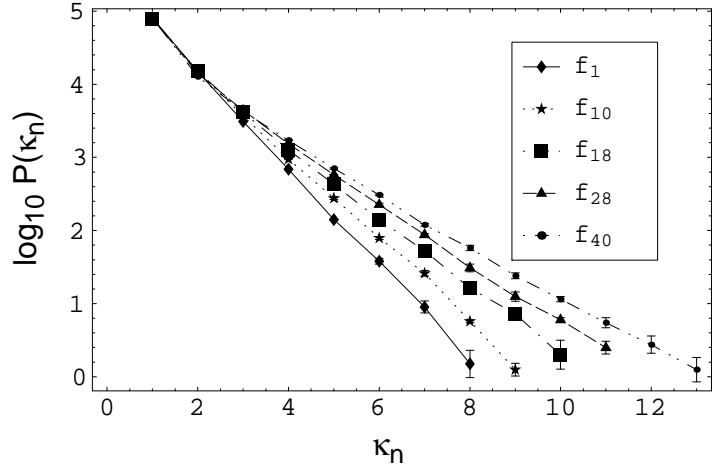}
\caption{Histogram of $\kappa_n$ for several $f_n$-avalanches.}
\label{fig:distr}
\end{figure}

We  measured $P_n(\kappa_n)$ during our simulations.  The logarithm of the
histogram of $\kappa_n$  for several values of $n$ is shown in Fig.\
\ref{fig:distr}.  It is clear that the behavior of $P_n(\kappa_n)$ is more complicated than
simple exponential decay. There is a short time fast decay and, for larger $n$, a long
time slow decay.  It is possible that the asymptotic behavior of $P_n(\kappa_n)$ for the
two largest values of $n$ in Fig.\
\ref{fig:distr}  both display exponential decay with the same value of $b$.
It is also possible that $b$ is a slowly decreasing function of $n$ or that the
asymptotic decay is slower than exponential.  If
$b$ decreases as a power of $n$ or if the tail of $P_n$ is a stretched exponential it will
still be the case that $\langle T(n_{\rm max}) \rangle$ is polynomial in $n_{\rm max}$
and thus polylogarithmic in $(f_c-\tf)$. On the other hand, if the tail of $P_n$ is a power
law then
$\langle T(n_{\rm max}) \rangle$ would be exponential in $n_{\rm max}$.  
The  data $P_n$ does not  suggest a power law tail
and the data for $\langle T(n_{\rm max}) \rangle$ appears to rule out exponential
growth however more numerical work is needed to fully understand the asymptotic
behavior of these quantities.

\section{Sequential \BS dynamics and P-completeness}
\label{sec:bspc} 
We have seen in the previous section that the construction of \BS histories can be
arranged to be carried out in parallel in a time that is polylog in the length of the
history.  There are however alternative dynamics for producing histories with the
correct statistical properties.  In this section we show that two of these dynamics have
the property of P-completenesss.  The property of P-completeness implies that it is not
possible to run these dynamics in parallel in polylog time.    The two dynamics considered here are the standard
dynamics by which the
\BS model is usually defined and a conditional dynamics where explicit values are not
assigned to sites until it is necessary.  Thus, if we knew only about
standard or conditional dynamics we would conclude that the \BS model generates
complex histories.  Parallel \BS dynamics shares features of both
standard dynamics and conditional dynamics. 

Although standard \BS dynamics cannot be implemented in parallel in polylog time, in
Sec.\ \ref{sec:parstan}   we show how to use the parallel construction of avalanches to 
achieve power law speed-up. 

\subsection{Standard \BS dynamics}
\label{sec:sbs}
\begin{figure}
\includegraphics{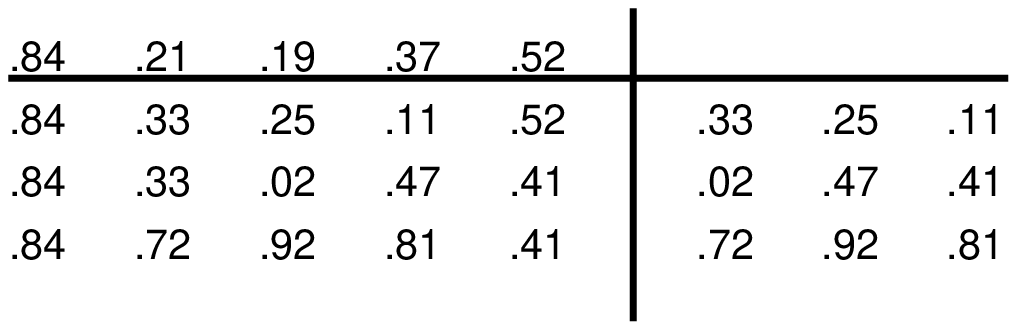}
\caption{An example of standard \BS dynamics.   The initial values are above the
horizontal line and time proceeds downward.  The numbers $y^{(n)}$ are to
the right of the vertical line.  During each time step, the extremal site and its two
neighbors are renewed by the triple of numbers to the right of the vertical line.}
\label{fig:standyn}
\end{figure}

Standard \BS dynamics formalizes the usual approach to generating \BS histories. 
Initially, the lattice is described by values $\{x_i(0)\}$.  A list of numbers
$\{y^{(n)} |\,  y^{(n)} \in [0,1] , n=1,2, \ldots , (2d+1) t_{\rm max}\}$  controls the
dynamics.  At each time step, the extremal site is located.  The extremal site and its
$2d$ neighbors are renewed using the next group of $2d+1$  $y$ values.  Periodic
boundary conditions are assumed.  Figure \ref{fig:standyn} shows an example of $d=1$
standard \BS  dynamics.  The numbers along the top row of the figure are $\{x_i(0)\}$
and the numbers in the three columns on the right are the groups
of $y$'s.  Successive states of the lattice are given in successive rows in the
lower left section of the figure. Each row is obtained by aligning the three $y$ values of
the row under the minimum $x$ value of the previous row. 

\subsection{P-completeness of standard dynamics}
P-completeness for a natural decision problem associated with standard \BS dynamics is
proved by a reduction from the circuit value problem (CVP).    

The inputs of the STANDARD BAK SNEPPEN decision problem
are the initial site values,
$\{x_i(0)\}$, the renewal numbers,  
$\{y^{(n)}\}$, a specified site $k$, time
$t_{\rm max}$ and bound $E$.  The problem is to determine whether $x_k(t_{max})<E$. 

In order to effect the reduction from CVP we need to show how to build logical gates
and wires that carry truth values.  We will show how to build AND gates, OR gates and
wires in the one-dimensional \BS model. This set of gates allows us to build monotone
circuits.  Since monotone CVP is  P-complete only for non-planar circuits, our
construction constitutes a P-completeness proof only after it is extended to $d>1$ \BS
models.  After we present the basic construction for $d=1$ we will discuss how to
extend the construction to $d>1$ and the prospects for proving
P-completeness for $d=1$.   

\begin{figure}
\includegraphics{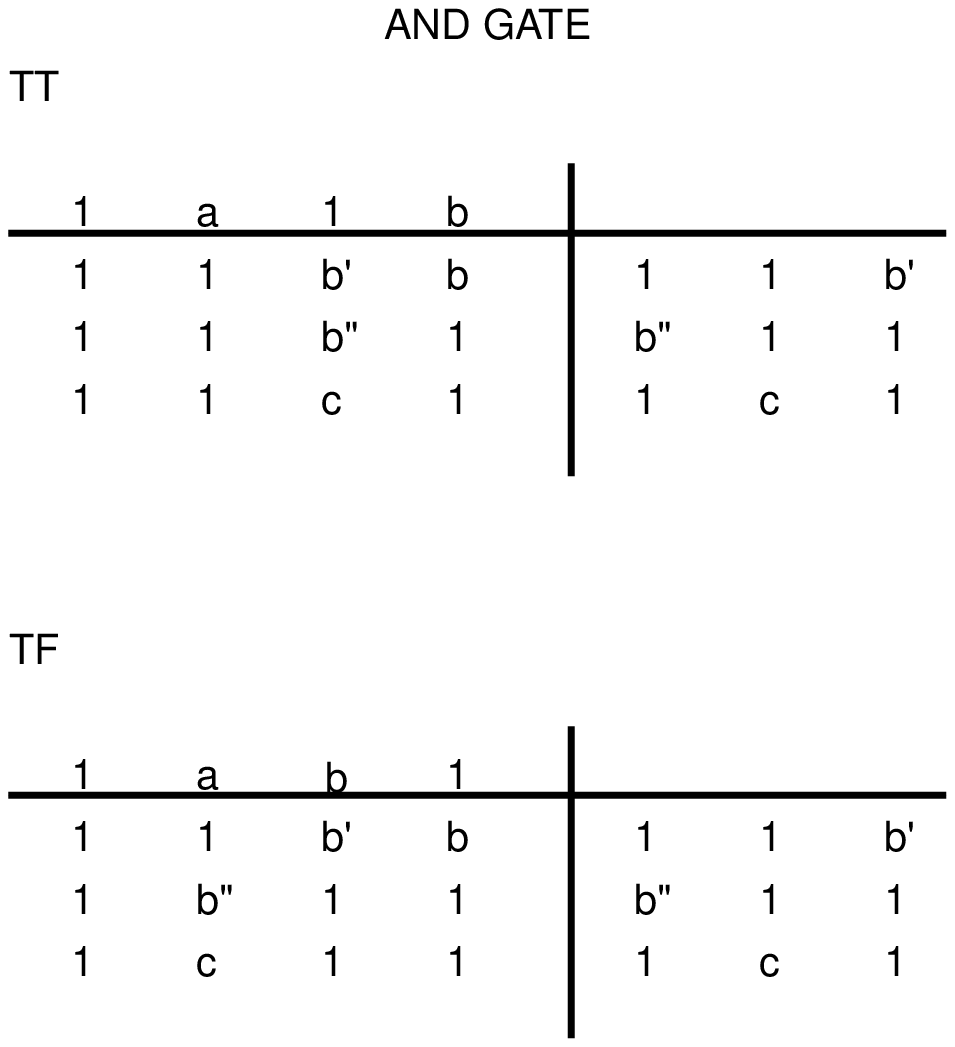}
\caption{Gadget for AND gate for standard dynamics.   The inputs $A$ and $B$ are
represented by the two pairs of sites above the horizontal line.    The output $C$ is
represented by the middle two sites  along the bottom row to the left of the vertical
line. The various numbers are ordered  $0<a<b<b^\prime<b^{\prime \prime}<c<1$.  The
top pictures shows the situation with both inputs are TRUE  yielding an output of TRUE.
The bottom picture shows the situation where the left input is TRUE and the right input
is FALSE yielding FALSE. }
\label{fig:stanand}
\end{figure}

In order to embed a circuit in standard dynamics, most of the sites of the lattice will
have the value 1.  A truth value is represented by a pair of neighboring sites $i$ and
$i+1$ such that within the pair, one site has the value one and the other has a value less
than one.  If the left member of the pair is a one the pair represents TRUE, otherwise it
represents FALSE.  A truth value is activated when its smaller member is the extremal
site of the lattice.  Once the truth value is activated, it can be moved by appropriate
choice of the triple of
$y$'s. The triple $(1,1,a)$ where $a$ is a suitable chosen number, $0<a<1$ will move a
truth value to the right while the triple
$(a,1,1)$ will move the truth value to the left.  AND gates are realized by
transporting two truth values next to one another.  
Suppose that initially, two truth values $A$ and $B$  are next to each other as shown in
the Fig.\ \ref{fig:stanand}.  The numbers
$a$, $b$, $b^\prime$, $b^{\prime \prime}$ and $c$ are ordered
$a<b<b^\prime<b^{\prime \prime}<c$ and, at the time the gate is activated, $c$ is less
than all the other numbers in the lattice.  The
top panel in Fig.\ \ref{fig:stanand} shows the situation where $A=$ TRUE and $B=$ TRUE.
Three time steps later,
$C$ represented by the middle two sites of the lattice is TRUE.  The
top panel in Fig.\ \ref{fig:stanand} shows the
situation where
$A=$ TRUE and $B=$ FALSE in which case $C=$ FALSE as desired.  The case $A=$ FALSE
and
$B=$ FALSE is the same as $A=B=$ TRUE except the states in the part of the lattice
representing the gate are shifted one lattice constant to the left so that $C=$ FALSE as
desired.  The final case $A=$ FALSE and $B=$ TRUE is left as an exercise for the reader.

\begin{figure}
\includegraphics{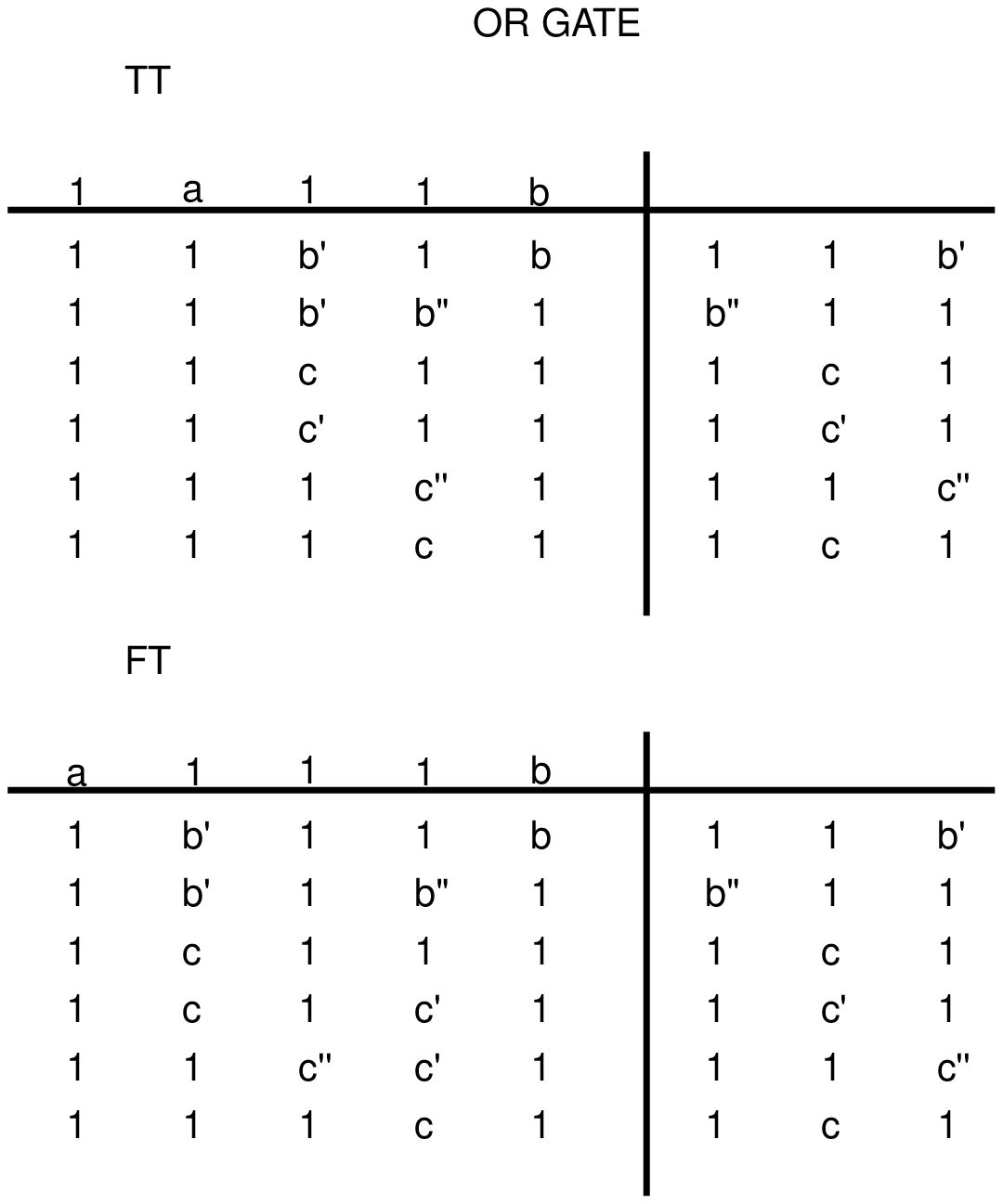}
\caption{Gadget for OR gate for standard dynamics.   The inputs $A$ and $B$ are
represented by the two outer pairs of sites above the horizontal line.    The output $C$ is
represented by the second and third sites to the left of the vertical
line. The various numbers are ordered  $0<a<b<b^\prime<b^{\prime
\prime}<c<c^\prime<c^{\prime\prime}<1$.  The top
pictures shows the situation with both inputs TRUE. The bottom picture shows the
situation where the left input is FALSE and the right input is TRUE, also yielding TRUE. }
\label{fig:stanor}
\end{figure}

OR gates are slightly more complicated than AND gates. Initially, the truth values are
brought together until they are separated by one site as shown in Fig.\ \ref{fig:stanor} 
The numbers are ordered $0<a<b<b^\prime<b^{\prime
\prime}<c<c^\prime<c^{\prime\prime}<1$ and, at the time the gate is activated,
$c^{\prime\prime}$ is less than all the other numbers in the lattice.  The top panel in
Fig.\ \ref{fig:stanor} shows the situation where
$A=$ TRUE and
$B=$ TRUE. Six time steps later, $C$, represented by the second and third columns to the
left of the vertical line, is TRUE.  The situation  $A=$FALSE and $B=$FALSE is seen to be
correct by translational invariance.  The case where $A=$ FALSE and
$B=$ TRUE is shown in the bottom panel of Fig.\ \ref{fig:stanor}.  Verifying the case $A=$
TRUE and
$B=$ FALSE is left as an exercise for the reader.

\begin{figure}
\includegraphics{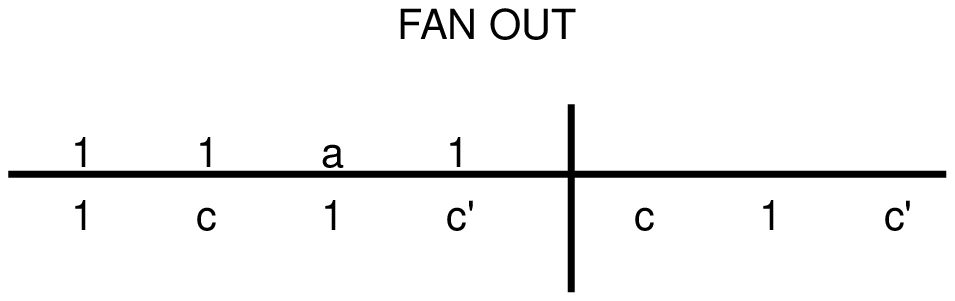}
\caption{Gadget for FAN OUT for standard dynamics.   The input $A$ is
represented by the middle pair of sites above the horizontal line.    The outputs $C$ and
$C^\prime$ are represented by the two pairs of sites to left of the vertical
line. The numbers are ordered  $a<c<c^\prime$. }
\label{fig:stanfan}
\end{figure}

We also need a gadget for fan-out.  Figure \ref{fig:stanfan} shows how two copies,
$C$ and
$C^{\prime}$ can be obtained from
$A$.  The figure shows the case where $A=$ TRUE.  The case where $A=$ FALSE is seen
to be correct by translational invariance.

Overall timing of the circuits is controlled by the numbers representing truth
values and used in gates.  Since, at each stage, the smallest number on the lattice is
selected, truth values represented by small numbers are activated first.  A truth value
$A$ that is active, meaning it is represented by a number $a$ that is the minimum
number on the lattice, can be put into storage for a pre-determined period by
choosing $y$ values $(1,b,1)$ with $b$ larger than all the values associated with logic
variables that are active until $A$ is active again.  In this way all the required $x$ and
$y$ values can be calculated with a fast parallel computation,  in advance knowing only
the activation time of each logic gate.  

AND and OR gates, fan-out, timing and wires (transporting truth values) are sufficient
for constructing arbitrary monotone Boolean circuits.  Since planar monotone circuits are
not P-complete, the preceding construction does not show that the $d=1$ STANDARD
BAK SNEPPEN problem is P-complete.  It is straightforward to construct all
the gadgets and arrange timings within $d>1$ standard \BS dynamics by simply padding
the set of $y$ values with 1's in the directions transverse to the planes in which the
action of the gadgets takes place.  Since truth values  can be transported  in any
direction, non-planar monotone circuits can be embedded in $d>1$ \BS systems.  The
conclusion is that the
$d>1$ STANDARD BAK SNEPPEN problem is P-complete.  Thus there can be no
exponential speed-up of standard dynamics through parallelism (unless it should turn
out that
$\PP=\NC$, a highly unlikely prospect).  Although the current proof does not yield a
P-completeness result for the planar
 STANDARD BAK SNEPPEN problem, we do not see a general reason
preventing the construction of non-monotone circuits and we cannot rule out the
possibility that the planar problem is also P-complete. 

\subsection{Power law speed-up of standard dynamics}
\label{sec:parstan}
At least for $d>1$, the P-completeness result precludes the possibility of solving the
STANDARD BAK SNEPPEN problem in polylog time.  Nonetheless we can achieve a
sublinear parallel time solution to the problem by invoking the
hierarchical construction of avalanches from parallel dynamics.  The history produced
by standard dynamics can be viewed as a sequence of
avalanches and these avalanches can be efficiently constructed
from the list of $y$ values using the parallel methods of Sec.\ \ref{sec:parf}.  The
road block in achieving exponential speed-up for standard dynamics does not lie in the
construction of avalanches, rather it is due to the presence of explicit initial values
instead of initial probabilities defined by $G(0)$ and made explicit only at the end of
construction.  The presence of explicit initial values means that the environment of each
avalanche is specified and the sequence of gap values cannot be determined in advance
of making the avalanches.  

The accelerated algorithm for the STANDARD BAK SNEPPEN problem proceeds as
follows.  The first gap value, $G(0)=\lm(0)$ is just the minimum among the initial
values.  The algorithm then calls a parallel avalanche subroutine to produce a
$G(0)$-avalanche from the initial sequence of $y$ values.  This avalanche (the full
avalanche, not just the backbone)  renews the sites that it covers while the rest of the
environment is unaffected.  At the ending time, $t_1$ of the $G(0)$-avalanche, the
extremal site is determined and its values determines the next gap value, $G(1)=f(t_1)$. 
Generally, suppose that the gap increases at times $t_k$, then $G(k)=f(t_k)$ and the
parallel avalanche subroutine is called to produce a $G(k)$-avalanche using the
remaining list of
$y$ values.  The origin of this avalanche is the extremal site $s(t_k)$ and the avalanche
renews all the sites it covers.  As the gap approaches the critical value, $f_c$, the
algorithm becomes increasingly efficient since larger and larger avalanches may be
made in parallel.

The parallel avalanche subroutine takes as inputs a gap value $G_k$ and the remaining 
 $y$ values,
 $\{y^{(n)} | \,  n=(2d+1)(t_k+1), \ldots , (2d+1) t_{\rm max}\}$ values.  Its output is the
$G_k$-avalanche that would have resulted if standard dynamics were used with the
same sequence of $y$ values.  The procedure is similar to that described in Sec.\
\ref{sec:parf} and uses the same hierarchy of
$f_n$-avalanches but care must be taken that the
$y$ values are used in the correct order.  Each block of $2d+1$ $y$ values is an
$f_0$-avalanche.  In parallel we must now group this sequence of $f_0$-avalanches into
$f_1$-avalanches.  The procedure for grouping avalanches is illustrated using Fig.\
\ref{fig:fseven}.  Each node on the bottom level of the tree  represents
blocks of $2d+1$
$y$ values.  In parallel we now independently start from each of these nodes and carry
out standard dynamics until an
$f_1$-avalanche is complete. For example, starting from bottom level node 2 it takes
two time steps to build an 
$f_1$-avalanche denoted by the 3 on the $f_1$ level of the tree but it only takes one
time step starting from bottom level node 3 since this $f_0$-avalanche is also an
$f_1$-avalanche.  The time that standard dynamics must be run to obtain each of the
$f_1$-avalanches defining the nodes on the $f_1$-level is just the maximum
in-degree of this level.  Thus far we have
made $f_1$-avalanches starting from each bottom level node, now we must determine
which of these avalanches to keep to reproduce the result of standard dynamics acting
on the
$y$ sequence.  The correct
$f_1$-avalanche to associate with each node in the
$f_1$-level is the one that includes all of the children of the node.  This choice insures
that all $y$ values are used.  For example, the
$f_1$-avalanche labeled 3 includes steps 2 and 3 rather than 3 alone.  

Having
assembled the correct set of
$f_1$-avalanches we can now group these into
$f_2$-avalanches and so on.  The new ingredient in going from the $f_{n-1}$ to
$f_n$-avalanches when
$n>1$ is that the generalized version of the standard dynamics is used as discussed in
Sec.\ \ref{sec:parf}.  As before, $G_k$-avalanches can be constructed in a time that is
polylog in the $S$, the duration of the avalanche.

The speed-up obtained due to the parallel avalanche subroutine can be estimated by the
scaling laws for the $\langle S \rangle$ and $f_c- \langle G(t) \rangle$.  Up to
logarithmic factors, it takes constant parallel time to produce a $G$-avalanche. 
Combining Eqs.\
\ref{eq:gamma} and
\ref{eq:gapt} we have
\begin{equation}
\frac{dT}{dt} \approx  \frac{1}{\langle S \rangle_G} \sim
(t/N)^{-\frac{\gamma}{\gamma-1}}
\end{equation}
where $T$ is parallel time and $t$ is sequential time.  Integrating the differential
equation yields, $T \sim t^{\frac{1}{\gamma-1}}$ so that power law speed-up is
obtained.

The important difference between parallel and standard dynamics is that in parallel
dynamics, sites that are not part of avalanches are specified probabilistically until the
last stage of the computation while, for standard dynamics, all sites are explicitly
determined on each time step.  This suggests the idea that an efficiently parallelizeable
dynamics might result from minimizing the number of explicitly determined
values until the last stage of the computation.  This idea lead to the conditional
dynamics described and analyzed in the next two sections.  

\subsection{Conditional \BS dynamics}

\label{sec:condprob}
In standard dynamics with the usual choice of initial conditions, lattice sites are assigned
random numbers uniformly distributed on $[0,1]$ and the lattice site with the smallest
number is selected for replacement.  In parallel dynamics only the backbone is
fully specified until the end of the computation. In  {\em conditional
dynamics} this idea is carried as far as possible.  Extremal space-time points are fully
specified and conditions are placed on all other space-time points until the end of the
calculation.   Conditional dynamics is similar to the Run Time Statistics approach in Ref.\
\onlinecite{FeCaGaPi}.

Initially, each site is equally likely to be extremal. A site
$s(0)$ is selected at random from among all
the sites of the lattice.  Suppose there are $N=L^d$ sites on the lattice. The first extremal
number,
$f(0)$ is chosen from the distribution of the minimum of $N$ numbers distributed
uniformly on
$[0,1]$.  Although we have not yet assigned numbers to
any other sites we now know that these sites are conditioned to be greater than
$f(0)$.  The next step in the standard \BS process is to renew 
$s(0)$ and its neighbors with numbers chosen on the interval $[0,1]$. Again, at this
stage, we do not need to explicitly know any of the numbers on the lattice
except to say that all the numbers on the lattice except $s(0)$ and its
neighbors are randomly chosen in the interval $[f(0),1]$ while $s(0)$ and its
neighbors are randomly chosen on the interval $[0,1]$.  We can characterize the
condition on site $i$ at time $t$ by the number $c_i(t)$ which indicates that the
random number at site $i$ is chosen from the uniform distribution on the
interval $[c_i(t),1]$.  We have that $c_{i}(1)=0$ for $s(0)$ and all its neighbors while
$c_{i}(1)=f(0)$ for all other sites.  The extremal site
$s(1)$ at
$t=1$ is a weighted random choice among the $N$ sites with site $i$ weighted by
$1-c_i(1)$.  The extremal value $f(1)$ is
chosen from the distribution of the minimum of $N$ numbers, where the $i^{\rm
th}$ number is chosen from the  uniform distributions on $[c_i(1),1]$.  The
minimum numbers
$f(0)$ and $f(1)$ now determine the conditions $c_i(2)$ and so on. 
Specifically, $s(t)$ and its neighbors are renewed at time $t+1$ so that 
$c_i(t+1)=0$ for $s(t)$ and its neighbors. For all other sites,
$c_i(t+1)= \max\{f(t) , c_i(t)\}$. 

One way to implement the selection of the extremal site at time $t$ given the
conditions $c_i(t)$ is to assign random numbers $r_i(t)$ on the interval
$[0,1]$ to each space-time point.  From these numbers, compute
$\tilde{r}_i(t)=c_i(t)+r_i(t)(1-c_i(t)) \equiv c_i(t) \xor r_i(t)$.  Note that $\tilde{r}_i(t)$ 
is uniformly distributed on $[c_i(t),1]$.  The extremal site $s(t)$ is the site
with the minimum among the $\tilde{r}$'s and $f(t)=\tilde{r}_{s(t)}(t)$.
Note that all the random numbers $r_i(t)$ can be assigned in advance.  It is this
realization of conditional \BS dynamics that is shown to be P-complete in Sec.\
\ref{sec:pc}.

After the conditional dynamics has run for a time $t_{max}$ we are left with the 
conditions
$\{c_i(t)\}$ together with a list of extremal sites $\{s(t)| t < t_{max}\}$ and
extremal values
$\{f(t)| t < t_{max}\}$.  From these data we can reconstruct a \BS history by
working backwards.  The approach is very similar to the construction of a full history
from the backbone in the parallel dynamics.  The final conditions,
$\{c_i(t_{max})\}$ can be used to obtain final values:
$\x_i(t_{max})$ are randomly chosen from the uniform distributions on
$[c_i(t_{max}), 1]$.   For each $i$, these final values are then the correct values
backwards in time through the latest time when site $i$ is renewed. 
Suppose site
$i$ is last renewed at time $t < t_{max}$ that is $i=s(t-1)$ or $i$ is a neighbor of
$s(t-1)$.  If
$i=s(t-1)$ then $\x_i(t-1)=f(t-1)$ and this value holds until the next earliest time that $i$
is renewed.  If site $i$ is renewed at time $t$ because it is the neighbor of the
extremal site $s(t-1)$ then $\x_i(t-1)$ is chosen from the uniform distribution on
$[\max(f(t-1),c_i(t-1)),1]$ and this value holds until the next earliest time that $i$ was
renewed.  Working backwards in this way a definite \BS history is constructed from
the set of extremal values, extremal sites and conditions.  It should be noted that
the reconstruction of a definite  history from the data generated by the conditional
dynamics can be carried out in parallel in polylog time.

During conditional dynamics, the values of the $c$'s become spatially
non-uniform however at times when the gap increases, almost all the $c$'s are
equalized.  If 
$G$ increases at time $t-1$  then for all $i$, $c_i(t)=G(t-1)$ except, of course, at the
extremal site
$s(t-1)$ and its neighbors where the $c$'s are reset to zero.

\subsection{P-completeness of conditional  dynamics}
\label{sec:pc}

P-completeness for conditional dynamics with $d \geq 1$ is proved by a reduction
from the monotone circuit value problem.  To carry out the reduction we need a way to
implement AND and OR gates and non-local fan-out.  We will explain how the reduction
works for the one-dimensional \BS model.  Since non-local fan-out can be implemented,
the P-completeness proof holds even for the one-dimensional problem. 

The inputs of the CONDITIONAL BAK SNEPPEN decision problem
are the initial conditions,
$\{c_i(0)\}$, the random numbers,  
$\{r_i(t)\}$, a specified site $k$, time
$t_{\rm max}$ and bound $E$.  The problem is to determine whether $c_k(t_{max})<E$. 

In the reduction from CVP, truth values
reside at specified sites and are represented by the values of
$c_i(t)$.    If
$c_i(t) <  \epsilon t $, the site has the value TRUE at time $t$ and if
$c_i(t) =1$ the site has the value FALSE. The number
$\epsilon$ is chosen to be small enough that the total time duration to evaluate
the circuit, $t_{\rm max}$ still yields small values, thus we set
$\epsilon t_{\rm max} < 1/4$.  Sites with truth values  are separated by background
sites where
$r_i(t)=c_i(t)=1$. If site $i$  has an initial truth value this is represented
by an initial value,
$c_i(0)=0$ or 1.  Sites that represent outputs of gates or fan-outs have
$c_i(0)=1$.

In addition to sites representing truth values, there is a single site called
$\F$ with values $r_{\F}(t)=\epsilon t$ and $c_{\F}(0)=0$.  A site $i$ carrying
a truth value can be {\em read} at time $t$ by setting $r_i(t)=0$.  If site $i$
is TRUE at time $t$ it is selected since $\tilde{r}_i(t)< \epsilon t$ whereas, if
$i$ is FALSE at time $t$ then $\tilde{r}_i(t)=1$ and $\F$ is selected at time
$t$.  

Finally, for each AND gate and each fan-out, we
need two neighboring auxiliary sites $\k$ and $\l$ with $c_\k(0)=0$
and $c_\l(0)=1$.

\begin{figure}
\includegraphics{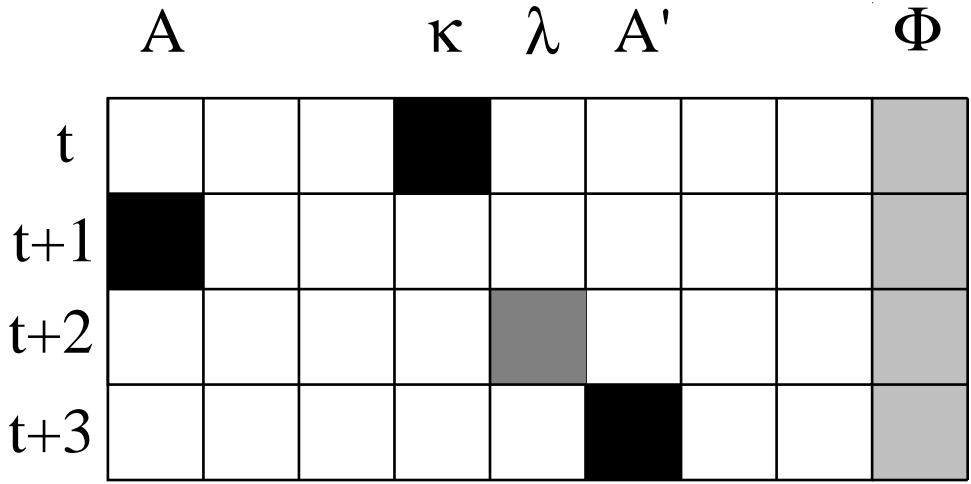}
\caption{Gadget for non-local fan-out of truth value at site A to site
A$^\prime$. }
\label{fig:fan}
\end{figure}

First, we show how to fan-out a truth value at site A at time $t$ to
a new site, A$^\prime$ at time $t+3$ using the gadget shown in Fig.\
\ref{fig:fan}.  In Fig.\
\ref{fig:fan}, white
squares have $r=1$, black squares have $r=0$ and gray squares
have small but non-zero $r$ values.  The gray squares in column $\F$ have 
$r_\F(t)=\epsilon t$ and the gray square in column $\l$ at time $t+2$ has
$r_\l(t+2)=3\epsilon /2$. The value of $c_A(t)$ depends on the truth value of A. 
If A is TRUE then $c_A(t) <   \epsilon t$ but if A is FALSE then
$c_A(t)=1$.  At time
$t$, site $\k$ is selected. First suppose A is TRUE.  Then, since
$c_A(t+1) < \epsilon t$ and $r_A(t+1)=0$ we have that site A is selected
at time $t+1$.  The selection of site A means that $c_\l(t+2) < \epsilon t$
and therefore, $\tilde{r}_\l(t+2) < \epsilon t + (3 \epsilon/2)(1-\epsilon t)
< (t+2) \epsilon$ so that site $\l$ is selected at time $t+2$. The selection of
$\l$ at time
$t+2$ insures that A$^\prime$ is set to TRUE for any time later than $t+2$ since
A$^\prime$ is refreshed at time
$t+3$ and $c_{{\rm A}^\prime}(t+3)=0$.  In Fig.\ \ref{fig:fan} we show
A$^\prime$ read at time $t+3$ but this is not necessary, it can
also read at any time after $t+3$. 

Now suppose A is FALSE.  Then site $\F$ is selected
at time $t+1$ and, consequently, $\tilde{r}_\l(t+2) = (t+1) \epsilon + (3
\epsilon /2)(1- (t+1)\epsilon) > (t+2) \epsilon$.  Thus site $\F$ is selected at
time $t+2$ and site A$^\prime$ fails to be refreshed so that $c_{{\rm
A}^\prime}(t+3)=1$ and A$^\prime$ is FALSE.

\begin{figure}
\includegraphics{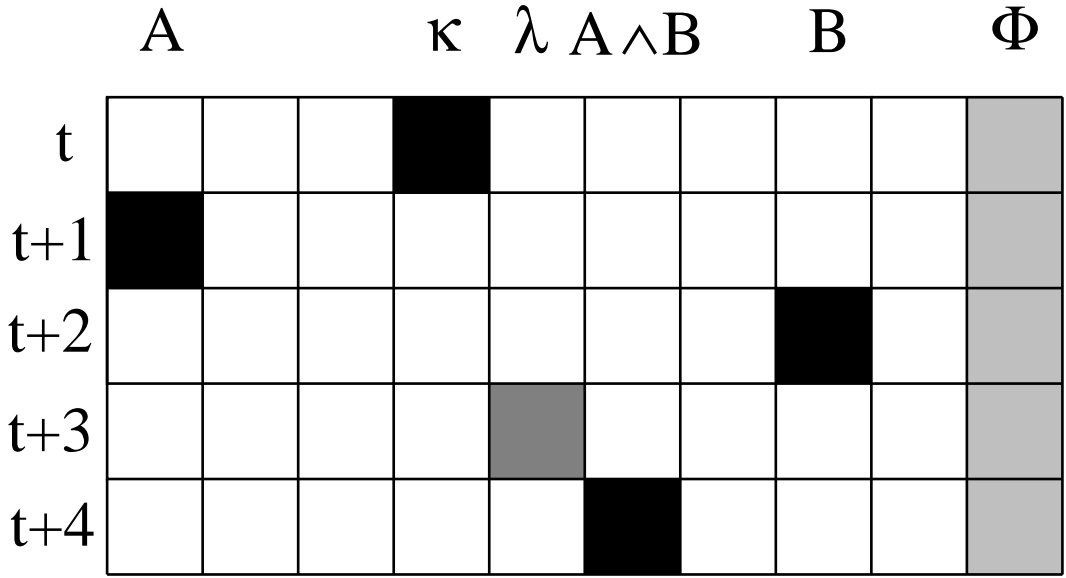}
\caption{Gadget for non-local AND gate.  Sites A and B may be anywhere in the
lattice. At time $t+4$, site A$\wedge$B is set to A AND B.}
\label{fig:and}
\end{figure}

The non-local fan-out just described can be easily extended to produce a
non-local AND gate as shown in Fig.\ \ref{fig:and}. As before, $r_\l(t+3)=3
\epsilon/2$.  The inputs A and B may be anywhere on the lattice, and the output
A$\wedge$B appears immediately to the right of the auxiliary sites $\k$ and
$\l$.  Suppose first that both A  and B are TRUE, then
$c_\l(t+3) < \epsilon t$ and $\tilde{r}_\l(t+3) < \epsilon t + (3
\epsilon/2)(1-\epsilon t) < (t+2) \epsilon$ so that $\l$ is selected at time
$t+3$ and A$\wedge$ B is refreshed at time $t+4$.  Thus the site A$\wedge$B is
correctly set to TRUE at time $t+4$.  On the other hand, suppose that A is FALSE
and B is TRUE.  Then
$\tilde{r}_\l(t+3)=(t+1) \epsilon + (3 \epsilon/2)(1-(t+1)\epsilon) > (t+3)
\epsilon$.  Thus site A$\wedge$B fails to be refreshed so that  A$\wedge$B is correctly
set to FALSE.  The other two possibilities for A and B are easily seen to work in the same
way.

\begin{figure}
\includegraphics{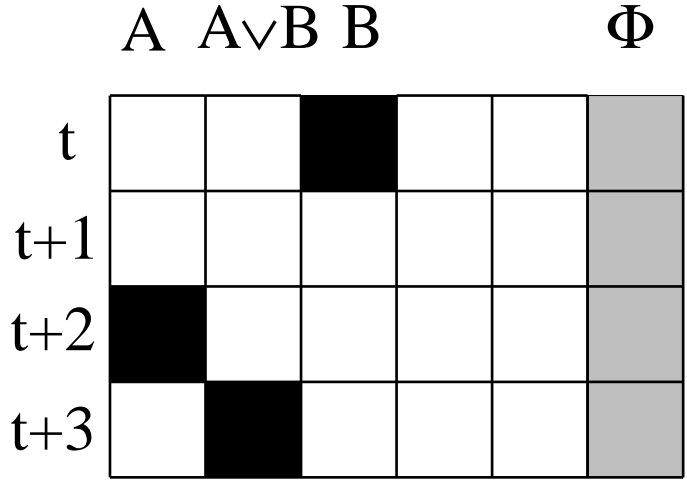}
\caption{Gadget for OR gate.  At time $t+3$, site A$\vee$B is set to A OR B.}
\label{fig:or}
\end{figure}

The OR gate is local in space but non-local in time and is shown in Fig.\
\ref{fig:or}.  Initially site A$\vee$B is set to FALSE however, at time
$t+2$, A$\vee$B is correctly set to A OR B since if either A or B are TRUE,
then A$\vee$B is refreshed and can be selected at any later time to be the input
to a fan-out or AND gate.  Figure \ref{fig:or} shows A and B read at time
$t$ and $t+2$ however any time separation between the reading A and B is
permitted, the only constraint on the OR gate is that both A and B must be read
before A$\vee$B is read.

These gadgets are sufficient to reduce monotone CVP to the one-dimensional
CONDITIONAL BAK SNEPPEN problem and show that the latter is P-complete.  Note that
the non-local character of the gadgets allows non-planar monotone CVP to be reduced to
the one-dimensional problem, in contrast to the situation for standard dynamics where
the P-completeness holds only for $d>1$.

\newpage
\section{Conclusions}
\label{sec:conc}

Our main result is that \BS histories can be efficiently generated in parallel. 
Specifically,  simulations and analytic arguments suggest that a history of length
$t_{\rm max}$ can be simulated on a PRAM with polynomially many processors in
average parallel time $\log^{\ord(1)} t_{\rm max}$ with the actual asymptotic behavior
close to $\log^2 t_{\rm max}$ . The exponential speed-up achieved by parallelization is
the result of re-organizing the history into a nested hierarchy of independent
avalanches.   The construction of a single
$\tf$-avalanche can be carried out in average parallel time $\log^{\ord(1)}
(f_c-\tf)$ which is exponentially less than the expected duration of the
avalanche
$\langle S
\rangle_{\tf} \sim (f_c - \tf)^{-\gamma}$.   

How is it possible to create long range correlations in space and time very quickly in
parallel?   First of all, it must be emphasized that the ground rules for parallel
computation with a PRAM allow for non-local transmission of information in a single
time step.  To see how space-time correlations can be set up in a parallel
time that is polylog in the correlation length or time, consider one step in the
hierarchical construction of avalanches.  For example, suppose two
$f_{n-1}$-avalanches are concatenated to yield a single $f_n$-avalanche.   The
correlation length and time are thereby increased by constant factors in a fixed number
of parallel steps.  Non-local transmission of information is needed to align the origin of
the second
$f_{n-1}$-avalanche with the final extremal site of the first $f_{n-1}$-avalanche
and it is this alignment of the two $f_{n-1}$-avalanches that increases the correlation
length and time.   The result is that spatio-temporal correlations grow exponentially in
the number of parallel steps.

In the parallel construction, an {\em independent} collection of $f_{n-1}$-avalanches are
concatenated to form an $f_n$-avalanche.  Nonetheless, \BS avalanches have a temporal
structure and exhibit aging~\cite{BoPa}.  Aging is consistent with the independence of
subavalanches for the following reason.  The last extremal value in the construction of
an $f_n$-avalanche must exceed $f_n$ and this extremal value is more likely to come
from the last $f_{n-1}$-avalanche used in the construction.   Thus, extremal values and
other properties of avalanches display aging.

Another example where critical correlations are set up much faster in parallel than
might be expected are cluster Monte Carlo algorithms~\cite{WaSw90,NeBa99} for critical
spin systems.  Each cluster sweep can be accomplished in polylog parallel time
on a PRAM with polynomially many processors using a parallel graph connectivity
algorithm.  The number of sweeps required to reach equilibrium scales as a small 
power (typically much less than unity) of the system size.  Thus long range critical
correlations are set up in a parallel time that is much less than the correlation length. 
This is quite different than physically realistic local dynamics where critical correlations
require a time that is at least quadratic in the correlation length.  

The intuition that \BS histories are generated by an inherently sequentially process is
not entirely wrong.  In fact, standard \BS dynamics was shown here to be
associated with a P-complete decision problem for $d>1$.  Standard  dynamics takes
as its inputs the initial values on the lattice and a sequence of $(2d+1)$-tuples of 
numbers that are used to renew the extremal site and its neighbors at each time step. 
The P-completeness result implies that it is almost certainly not possible to parallelize
standard dynamics so that the resulting history is generated in polylog time.  The key
problem that prevents full parallelization of standard dynamics is that all site values are
explicitly defined at every step in the construction. In contrast, parallel dynamics does
not assign explicit values except in the avalanche backbones until the end of the
calculation.  For the \BS model, predicting the future starting from explicit initial
conditions is harder than sampling a typical history.

 Conditional dynamics is a third
method for generating \BS histories that has the smallest set of space-time sites
explicitly defined until the end of the calculation.  Nonetheless, conditional dynamics is
also associated with a P-complete decision problem and so cannot be efficiently
parallelized.  Parallel dynamics shares features in common with both conditional and
standard dynamics and is intermediate in the degree to which it avoids explicit
specification of site values until the end of the computation.  

The results presented here reflect characteristics of the \BS model that are independent
of the PRAM model of computation in which they were presented.  The primary result of
the paper can equivalently be stated in terms of families of Boolean circuits. 
Specifically, we can envision a hard-wired device composed of logical gates and random
bit generators.  The gates are wired in a feedforward direction so that, in  generating a
history, each gate is used only once.  When the circuit is activated it produces a
statistically correct
\BS history.  For each system size
$L$ and time $t_{\rm max}$ we need a different circuit.  Because of the equivalence of
PRAMs and circuit families, our result can be stated in terms of the logical depth of the
circuit. Logical depth is the longest sequence of gates between any of the random bit
generators and the output representing the history.  The existence of an efficient
parallel algorithm implies that there is a uniform family of the circuits for generating
\BS histories whose depth scales polylogarithmically in $t_{\rm max}$, independent of
$L$.  The number of gates in the circuit is bounded by a power of $t_{\rm max}L$.  The 
actual running time for any real circuit generating \BS histories would, of course, be
polynomial and not polylogarithmic in $t_{\rm max}$ because of the need for wires
connecting distant gates required for establishing the long range spatio-temporal
correlations.

A variety of non-equilibrium models in statistical physics  have
been employed to study the spontaneous emergence of complexity.  Besides the \BS
model, other examples include diffusion limited aggregation~\cite{WiSa} (DLA), invasion
percolation and sandpiles.  Besides their application to specific physical phenomena, 
these models have a broad appeal because they are governed by simple microscopic
rules and yet they display  self-organized criticality.  It has been argued
that they perhaps shed light on far more complex phenomena found, for example, in 
biology or economics.  Surely, however,  biological and economic systems generate
histories that have polynomial rather than polylogarithmic logical depth.   Perhaps
one criterion for a model of the spontaneous emergence of complexity is that it
should generate histories that require more than polylogarithmic depth to simulate. 
Neither the \BS model nor invasion percolation~\cite{MaGr} satisfy this requirement.  On
the other hand, both diffusion limited aggregation and the Bak, Tang and Wiesenfeld
(BTW) sandpile model~\cite{BaTaWi}  are related to P-complete
problems~\cite{MaGr96,MoNi} which strongly suggests that the clusters or avalanches
associated with these models cannot be simulated in polylog depth.   However,
neither model generates histories much deeper than a power of the system size.  In the
case of DLA, the cluster size cannot exceed the system size so
that the length of the history is bounded by $t_{\rm max}  \leq L^D$ where $D$ is the
fractal dimension of the cluster.  Sandpile models are well-defined for arbitrarily long
times. However, the BTW model has an Abelian property, which allows avalanches to be
combined in any order.  Thus, it is possible to rearrange long histories so they are
generated in polylog parallel time.  From a given initial condition, it requires 
$\ord(L^{d+2})$ steps to relax the system~\cite{MoNi}. A sequence of avalanches due to
randomly dropping sand on single sites can be organized into a binary tree using the
same idea as adding numbers in parallel.    The result is that a history of length $t_{\rm
max}$ can be simulated in parallel in time $\ord(L^{d+2} \log t_{\rm max})$.  By
contrast,  we find that \BS histories can be simulated in parallel time
$\log^{\ord(1)} t_{\rm max}$ independent of system size.

\section{Acknowledgements}
This work was supported by NSF grants DMR-9978233.  We thank Ray Greenlaw, Cris
Moore and Stefan Boettcher for useful discussions.

\end{document}